# Evaluation of hybrid run-time power models for the ARM big.LITTLE architecture

Krastin Nikov, Jose L. Nunez-Yanez
University of Bristol
Bristol, UK
e-mail: {kris.nikov, j.l.nunez-yanez}@bristol.ac.uk

Matthew Horsnell
ARM Ltd.
Cambridge, UK
e-mail: matt.horsnell@arm.com

*Abstract*—Heterogeneous processors, formed by binary compatible CPU cores with different microarchitectures, enable energy reductions by better matching processing capabilities and software application requirements. This new hardware platform requires novel techniques to manage power and energy to fully utilize its capabilities, particularly regarding the mapping of workloads to appropriate cores. In this paper we validate relevant published work related to power modelling for heterogeneous systems and propose a new approach for developing run-time power models that uses a hybrid set of physical predictors, performance events and CPU state information. We demonstrate the accuracy of this approach compared with the state-of-the-art and its applicability to energy aware scheduling. Our results are obtained on a commercially available platform built around the Samsung Exynos 5 Octa SoC, which features the ARM big.LITTLE heterogeneous architecture.

## I. INTRODUCTION

Mobile computing is the fastest growing consumer technology area in recent history and it is a major part of the semiconductor industry. However recent trends like the slowdown of Moore's law and the emergence of Dark Silicon [1] make it harder and harder for industry to keep up with consumer demand in terms of performance and energy efficiency. It is becoming ever more critical to develop better energy management solutions for the current and future generations of computing platforms. The increased sophistication of these platforms with heterogeneous processors such as ARM big.LITTLE [2] makes this a more challenging task. For example, "race-to-idle" policies which minimize CPU work time have been found to be the best way to optimize power consumption for most desktop CPUs (Intel, AMD). However Imes et al. [3] shows that in contrast to Desktop Systems, Embedded Systems (ARM) achieve better power efficiency with a "never-idle" scheduling strategy. In the context of the ARM big.LITTLE System-on-Chip (SoC), the Energy Management system has an additional level of complexity. For that purpose ARM and the supporting community have developed patches for the Linux and Android Operating Systems, which support a custom scheduler for big.LITTLE [4]. The scheduler is a natural extension of Dynamic Voltage and Frequency Scaling (DVFS), which allows tasks to be migrated from one CPU cluster to the other. Thanks to the Cache Coherent Interconnect (CCI) the overhead of migrating the task is kept low. The scheduler relies on migration thresholds to decide when it is time to migrate the task to a performance or a power-efficient CPU. The threshold levels are dependent on implementation, but in all cases are chosen far enough apart to prevent overzealous switching. With the current movement towards improving energy efficiency for mobile systems, significant improvements can be made by developing platform-specific solutions which can exploit system capabilities at a lower level.

An example of a more advanced scheduling strategy is using power models to intelligently migrate tasks between the heterogeneous cores. In this paper we present our efforts to develop a run-time Power Model, which could be used to improve the existing scheduler available for big.LITTLE. We have explored, tested and improved published power model generation methodologies on a big.LITTLE development platform. Our focus is real-time usability in a dynamic system. We also investigate the tradeoff between model complexity/accuracy and have developed a hybrid approach.

The novel contributions of this paper can be summarized as follows:

- We build configurable run-time CPU power models using 3 components: physical information of the development platform, Performance Monitoring Unit (PMU) events and CPU state information.
- We evaluate our approach against a purely physical model (Takouna [5]), PMU based model (Pricopi [6]) and CPU state based model (Walker [7]) and we show improved capability to predict power when tested on cBench using a big.LITTLE development platform.
- We have identified that tuning the model for every CPU frequency level is much more accurate than having a unified model.
- The developed models can be configured to only include those component available to the model system so they can be easily reconfigured for other big.LITTLE development platforms that for example might not have access to PMU events or CPU states.
- We explore the cross-prediction capabilities of our models to predict runtime CPU power using information from the other CPU core-type available on the Heterogeneous SoC. These models could be used to guide a scheduling algorithm in migrating tasks efficiently.

## II. RELATED WORK

The ability to model system and CPU power has been investigated ever since the very birth of the semiconductor industry [5], [8], [9]. In this section we summarize some of the published methodologies for developing power models and contrast them to our approach.

Modern power models use high-level events to derive power, since they are much easier to observe. Takouna et al. [5] present a very minimal linear power model for the Intel Xeon E5540 CPU, which uses frequency and number of cores to predict power with an average of 7% error. A limitation to this approach is that it cannot predict the changes in power consumption at a particular frequency level, as it will only model the average power for that frequency.

A successful way to observe finer changes in program execution and power is to use hardware system information available from the Performance Monitoring Units (PMU) on a CPU. Historically PMUs have been used to estimate performance, but researchers have been successful in also estimating CPU/system power consumption using PMU hardware events. Bertran et al. [9] use PMU events to compute accurate fine-grained power models. They present an empirical 2-level Functional Level Power Analysis (FLPA) model for an Intel Core2 Duo CPU and are able to successfully predict program phases with 83.91% success. In contrast to this is the work of Nunez-Yanez et al. [8] makes a strong case that system-level modelling is better that component level modelling, They use a large number of PMU events collected with a simulator on an ARM Cortex-A9, to train a linear model using linear regression. Instead of using micro benchmarks they use cBench [10] as a workload stress the entire system as a whole and report an average of 5% estimation error.

Pricopi et al. [6] develop complex models for predicting performance on big.LITTLE SoC by predicting the Cycles Per Instruction (CPI) stack. As part of their work, however, they have also built an accurate empirical model for the big.LITTLE ARM Cortex-A15 CPU using 7 PMU events. They train the model with an average error of 2.6% when trained and tested on SPEC CPU2000 and SPEC CPU2006 benchmark suites. One of the main points they argue is that the ARM Cortex-A7, the other CPU in the big.LITTLE SoC, exhibits very little power variation, so they choose not to build a complicated model for it.

Another approach to predicting power is using information about the CPU state and utilization. Walker et al. [7] present a CPU frequency and utilization based model for a different big.LITTLE platform, which did not have support for PMU. They obtain information about CPU time spent in idle using information available from the Linux kernel running on the device. Tested on the same workload as the PMU model, the CPU frequency and idle time model achieves 10.4% and 8.5% error for the ARM Cortex-A7 and ARM Cortex-A15 respectfully.

### III. METHODOLOGY

#### A. Development platform and workload

The development platform is based around an ARM big.LITTLE SoC developed by Samsung. The system combines high performance CPU cores with power efficient ones in a configurable combination. The two core types use the same instruction set architecture (ISA) so are able to execute the same compiled code. The aim is to achieve better power efficiency by using the heterogeneity of the system to direct tasks towards the core type they are better suited to.

The ODROID XU3[11] development platform by Hardkernel, features an implementation of the Samsung Exynos 5422 SoC, similar to the SoC used in the Galaxy S5. The SoC utilizes ARM's big.LITTLE technology and comprises four ARM Cortex-A15 (big) [12] and four ARM Cortex-A7 CPU (LITTLE) [13]. The board also has four sensors, measuring the A15, A7, RAM and GPU power, current and voltage. The platform was set up with a Lubuntu 12.04 running odroidxu3-3.10.y kernel available from the board support team.

The ODROID XU3 has a broad DVFS range with the Cortex-A15 having 18 available frequency levels ranging from 0.2-2 GHz and 5 corresponding voltage levels and the Cortex-A7 with 14 available frequency levels from 0.2-1.4 GHz with 5 available voltage levels.

The ideal workloads for this system would be exhaustive benchmarks with diverse behavior in order to capture different scenarios and long runtime, which is why we used cBench [10].

Fig. 1 shows the CPU energy/frequency relationship. The convex curves show that the lowest energy point is not simply the smallest voltage/frequency level and therefore is not always predictable without knowing the workload. This further supports our claim that accurate power models could be extremely useful for dynamic energy management.

We have observed a significant range in power consumption for both CPU core types while running the workload. We recorded an average power variation of up to 126% at 1.6 GHz and 143% at 0.8 GHz for the Cortex-A15 and Cortex-A7 respectively. This seems to indicate that the conclusions made by Pricopi et al. [6], which state that due to its simplicity, the Cortex-A7 power can be modeled by using single average number for each frequency level, must be revisited. We argue that due to the high variation of the Cortex-A7 a more complex model should be used to accurately predict the Cortex-A7's power consumption.

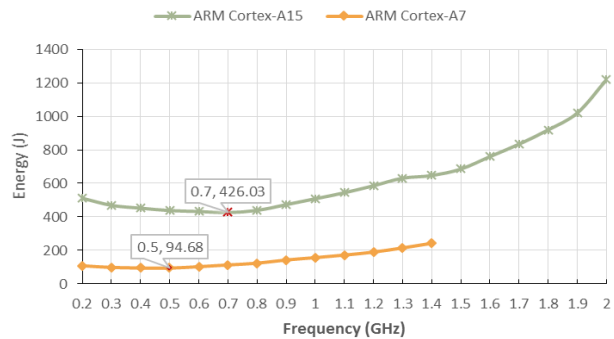

Figure 1. Energy/Frequency relationship on ODROID XU3

#### B. Model generation and verification

After we collect the results from the train set we use linear regression in the form of Ordinary Least Squares (OLS) to compute the predictor equation

$$\beta = (X^T X)^{-1} X^T y = (\sum x_i x_i^T)^{-1} (\sum x_i y_i).$$

Power is used as the dependent variable $\beta$ in the above equation, also known as regressand. The events are expressed as the *X* vector of independent variables, a.k.a. regressors. The OLS method outputs a vector $\beta$, which holds coefficients extracted from the activity vectors. Then the equation:

$$P_{CPU} = \beta_0 + (\beta_1 \cdot event_1) + (\beta_2 \cdot event_2) + \cdots (\beta_n \cdot event_n)$$

is used to estimate power usage using a new set of events. We evaluate the accuracy of the modelled equation by using a test set with a new set of power values and events and measure the percent difference (error) between the measured power and the estimated power by plugging the new events in the equation.

In general machine learning approaches like this are quite dependent on the inputs and equations used. If the model is too simple it might not give accurate predictions, because it does not use a sufficient number of characteristics (events/regressors) to fit the data properly. On the other hand a too complicated model, using many events, can make it hard to compute in real-time and can be prone to overfitting the training data and if the training data is not broad enough it might perform poorly on future types of work that have not been included in the training data. There is a fine balance between simplicity/real-time usability and good performance, but there is a lot of evidence that linear regression models can be used in power optimization techniques in embedded systems and produce accurate models [8], [14], [15].

The models are built from obtained measurements from the board using the OLS algorithm. Correlation analysis is done for events on the train set and only the ones that best model power (minimal error) are used in the final model.

The model we have developed consists of 3 distinct components:
1. Physical, expressed in (1), which has physically controlled regressands such as CPU voltage, CPU frequency and CPU temperature. CPU frequency is proven to be highly correlated to power consumption [5] however we believe, that voltage and temperature also play a crucial role in determining dynamic power and are not dependent on frequency. We show that adding them increases model accuracy, compared to just using frequency.
2. PMU events, expressed in (2), which has events available for both A15 and A7. We use PMU events common to both CPUs, instead of ones available for only the Cortex-A15. This is necessary if we want to use the power model to help make scheduling decision, since we need to have the same environments as inputs to both models in order to decide where we want to migrate the tasks. We consider 10 common PMU events available to both CPU core types and after mathematical analysis choose the top 4 most correlated to power plus CPU cycle count, since that is the maximum number of events the Cortex-A7 PMU can read concurrently without multiplexing. The list of PMU events is available in the Technical Reference Manuals available for both CPUs [12], [13].
3. CPU state, expressed in (3), uses information about time spent in different relevant CPU states and is expanding on the work done by Walker et al. [7] by not just using CPU Idle but also other states. In our work we have concluded that just using CPU Idle does not capture the dynamic CPU power accurately since during our runs of the workload the CPU spent very little time in that state.

$$P(W) = const. + (\alpha_1 \cdot CPU\ Voltage) + (\alpha_2 \cdot CPU\ Frequency) + (\alpha_3 \cdot CPU\ Temperature) \quad (1)$$

$$P(W) = const. + (\alpha_1 \cdot Cycles) + (\alpha_2 \cdot L1\ D\ Access) + (\alpha_3 \cdot L1\ I\ Access) + (\alpha_4 \cdot Instructions) + (\alpha_5 \cdot Mem\ Access) \quad (2)$$

$$P(W) = const. + (\alpha_1 \cdot CPU\ User) + (\alpha_2 \cdot CPU\ System) + (\alpha_3 \cdot CPU\ Idle) + (\alpha_4 \cdot CPU\ IO\ Wait) + (\alpha_5 \cdot CPU\ IRQ) + (\alpha_6 \cdot CPU\ Soft\ IRQ) \quad (3)$$

We then combine the events from the 3 component models listed above to develop a Physical + PMU events (called P2) model, shown in , (4) and a model using all 3 groups of events Physical + PMU events + CPU state (called P2S) model, shown in (5). The P2 and P2S model achieve the high baseline accuracy resulting from the use of the physical events with the ability to follow the power usage during program execution, thanks to the PMU events and CPU state information.

$$P(W) = const. + (\alpha_1 \cdot CPU\ Voltage) + (\alpha_2 \cdot CPU\ Frequency) + (\alpha_3 \cdot CPU\ Temperature) + (\alpha_4 \cdot Cycles) + (\alpha_5 \cdot L1\ D\ Access) + (\alpha_6 \cdot L1\ I\ Access) + (\alpha_7 \cdot Instructions) + (\alpha_8 \cdot Mem\ Access) \quad (4)$$

$$P(W) = const. + (\alpha_1 \cdot CPU\ Voltage) + (\alpha_2 \cdot CPU\ Frequency) + (\alpha_3 \cdot CPU\ Temperature) + (\alpha_4 \cdot Cycles) + (\alpha_5 \cdot L1\ D\ Access) + (\alpha_6 \cdot L1\ I\ Access) + (\alpha_7 \cdot Instr.) + (\alpha_8 \cdot Mem\ Access) + (\alpha_9 \cdot CPU\ User) + (\alpha_{10} \cdot CPU\ System) + (\alpha_{11} \cdot CPU\ Idle) + (\alpha_{12} \cdot CPU\ IO\ Wait) + (\alpha_{13} \cdot CPU\ IRQ) + (\alpha_{14} \cdot CPU\ Soft\ IRQ) \quad (5)$$

The developed composite models P2 and P2S are compared to relevant published work. Equation (6) represent the single-thread interpretation of the model presented by Takouna et al. [5] from the University of Potsdam. We call it UoP model for short.

$$P(W) = const. + (\alpha_1 \cdot CPU\ Frequency) + (\alpha_2 \cdot CPU\ Frequency^2) \quad (6)$$

Walker et al. [7] from University of Southampton present 2 models built on big.LITTLE using CPU idle time, presented in (7) and (8). Equation 8 has added CPU frequency related regressands to help capture the entire frequency range they report around 10% accuracy on the A15 and A7. We call them UoS CPU Idle model and UoS full model for short.

$$P(W) = const. + (\alpha_1 \cdot CPU\ Idle) + (\alpha_2 \cdot CPU\ Idle^2) \quad (7)$$

$$P(W) = const. + (\alpha_1 \cdot CPU\ Idle) + (\alpha_2 \cdot CPU\ Frequency) + (\alpha_3 \cdot CPU\ Idle \cdot CPU\ Frequency) + (\alpha_4 \cdot CPU\ Idle^2) + (\alpha_5 \cdot CPU\ Idle^2 \cdot CPU\ Frequency) \quad (8)$$

Pricopi et al. [6] from Cambridge Silicon Radio present a PMU based model built for the ARM Cortex-A15 running at 1Ghz. We call this model CSR for short and the events they use are shown in (9), where N indicates total number of instructions, INT - number of integer instructions and VFP - floating point instructions. They report around 2.6% average error.

$$P(W) = const. + (\alpha_1 \cdot IPC) + \left(\alpha_2 \cdot \frac{INT}{N}\right) + \left(\alpha_3 \cdot \frac{VFP}{N}\right) + \left(\alpha_4 \cdot \frac{L1\ D\ Access}{N}\right) + \left(\alpha_5 \cdot \frac{L2\ Access}{N}\right) + \left(\alpha_6 \cdot \frac{L2\ Refill}{N}\right) \quad (9)$$

Finally (10) shows the CSR model extended with physical and CPU state information as done in our own approach. The CSR model accesses more PMU events since it is designed exclusively for the A15 while P2 and P2S work for both A7 and A15. Equation 10 will serve to illustrate how adding physical and CPU state information can greatly improve performance.

$$P(W) = const. + (\alpha_1 \cdot CPU\ Voltage) + (\alpha_2 \cdot CPU\ Frequency) + (\alpha_3 \cdot CPU\ Temperature) + (\alpha_4 \cdot IPC) + \left(\alpha_5 \cdot \frac{INT}{N}\right)\left(+\alpha_6 \cdot \frac{VFP}{N}\right)\left(+\alpha_7 \cdot \frac{L1\ D\ Access}{N}\right) + \left(\alpha_8 \cdot \frac{L2\ Access}{N}\right) + \left(\alpha_9 \cdot \frac{L2\ Refill}{N}\right) + (\alpha_{10} \cdot CPU\ User) + (\alpha_{11} \cdot CPU\ System) + (\alpha_{12} \cdot CPU\ Idle) + (\alpha_{13} \cdot CPU\ IO\ Wait) + (\alpha_{14} \cdot CPU\ IRQ) + (\alpha_{15} \cdot CPU\ Soft\ IRQ) \quad (10)$$

## IV. MODEL ACCURACY ANALYSIS

In this section we present the model results in graphical form. We present 2 types of models – unified single-equation model trained over points on the entire frequency range of the CPU and models trained on a per-frequency basis. We show that the per-frequency models perform significantly better than the full frequency range models, which to our knowledge is an observation not previously documented on big.LITTLE, since researchers have focused largely on unified models.

### A. Full frequency range models

We start first highlighting the results using the models from our own methodology and show the performance of the individual component models as well as the grouped P2 and P2S models. All these models are trained on the train set using all points available from all frequencies and then tested on all the points of the test set. The percent error between the predicted and measured power is calculated and the averages for each frequency level are presented in Fig. 2 for the ARM Cortex-A15 and Fig. 3 for the ARM Cortex A7. Our observations show that the resulting complex P2S model is not always the best performer, instead simpler models just using physical characteristics like CPU frequency, voltage and temperature tend to have the lowest average error, so in general the CPU power consumption is mostly influenced by the physical information and this is not surprising. However introducing PMU event information and CPU state information does not improve performance as much as we expected. This is due to the fact that both PMU events and CPU state information in general have quite high variance for different frequencies (especially low and high frequencies, where we find increase in error compared to the mid-range frequencies), so instead of successfully predicting the finer power changes during workload execution they introduce larger margin of error. This is due to the fact that our sampling rate is time based and the number of samples and the event values at each sample point would vary from frequency to frequency. Another thing to note is that we expected the models for the ARM Cortex-A7 to perform a lot better, since it is a much simple CPU in terms of architecture, however this does not seem to be the case.

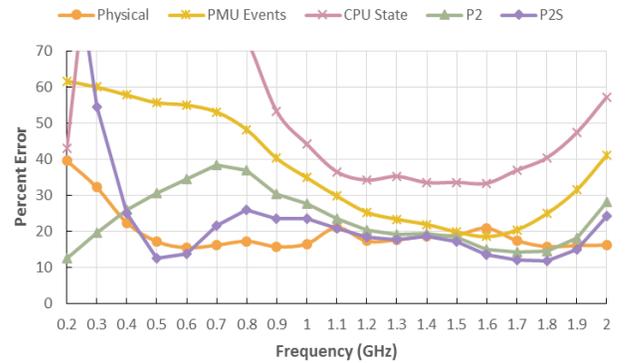

Figure 2. Full frequency models on Cortex-A15

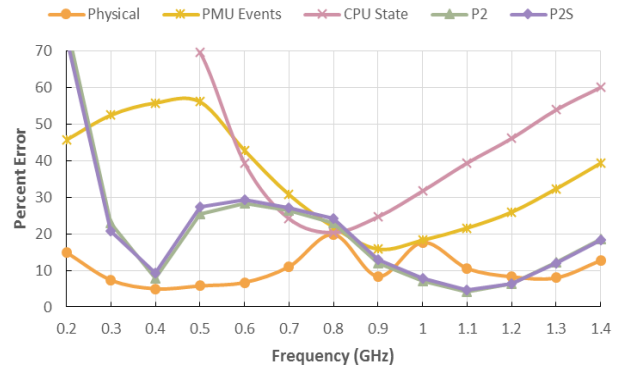

Figure 3. Full frequency models on Cortex-A7

Our results shown on Fig. 2 and 3 indicate that the best full frequency model for the Cortex-A15 and the Cortex-A7 is Physical with an average error of 19.95% and 10.46% respectively.

We observe that the accuracy of the models is affected by the CPU frequency and most models achieve the highest accuracy on the mid-range of frequency levels, 1-1.6 GHz for the Cortex-A15 and 0.6-1.2 GHz for the Cortex-A7. Ideally we would like a flat line close to zero so accuracy is high and invariable with frequency. The fact that new frequencies involve different voltages means that it is challenging for the model to maintain accuracy across all the whole frequency range.

*B. Per-frequency models*

Based on previous observations on the variability of PMU events and CPU state information between frequency levels we decided to try and calculate the models on a per-frequency basis. We decided not to change any of the equations, despite the frequency and voltage level remaining the same for the training, since we just use all the data point at one voltage/frequency level. This is because physical information gives a very good prediction for average power. Fig. 4 and 5 show that our analysis is justified by the results. We see a large reduction for the average model error and that adding CPU state and PMU event predictors improve accuracy significantly compared to just using the Physical model.

Overall we report an average of 8% error for the ARM Cortex-A15 and 5% error for the per frequency level P2S power model, which is on our target goal for the Cortex-A7. The fact that the model error is not a straight line means that there is some relationship we don't accurately cover. It could be possible to add mechanistic information (e.g. pipeline length, microarchitecture details) to try to improve accuracy further however this could translate in an excessively complex model, not suitable for real-time scheduling.

## V. EXPERIMENTAL EVALUATION

This section describes our results after testing some the published models described in section 3.2 using our workflow on the ODROID XU3. Fig. 6 and 7 show them compared to our best performing per-frequency model P2S for the Cortex-A15 and Cortex-A7 respectively. In both cases the P2S model performs a lot better with the exception of the Updated CSR model enhanced by us.

This is to be expected since the Updated CSR is a more complex model with larger, more specialized list of PMU events that are used in the model. The similarity between the "UoS full" and "UoP" models highlights that using CPU idle as a regressand does not significantly improve performance over the physical regressands that the UoP model uses. However including the other CPU states as regressands improves accuracy albeit only marginally and just on the Cortex-A7. Notice that Fig. 7 does not include results for the CSR models because a model has not been developed for the Cortex A7 in their approach.

This shows that our work is effective in producing models with competitive accuracy to the state-of-the art in current research.

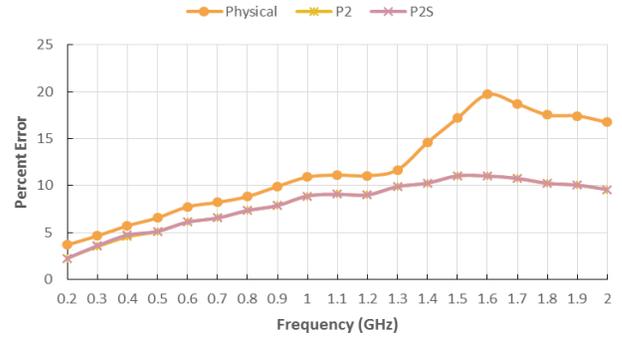

Figure 4.   Per-frequency models on Cortex-A15

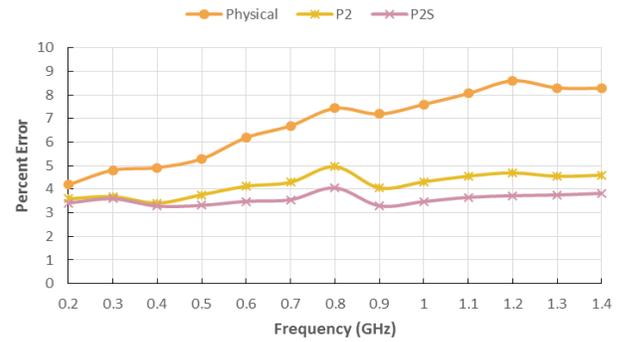

Figure 5.   Per-frequency models on Cortex-A7

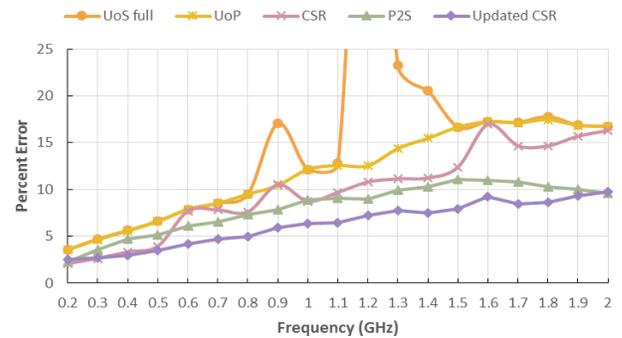

Figure 6.   Model comparisson on the Cortex-A15

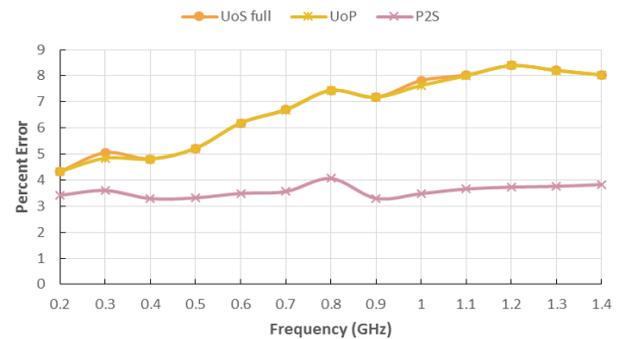

Figure 7.   Comparison between P2S and other published models on the Cortex-A7

## VI. INTER-CORE MODELS

We have also explored the usability of the developed P2S model to accurately predict inter-core power. This is done in order to see if the modelling methodology can be used to aid scheduling decisions for migrating a task from one core type to the other. We have tested 2 variants of the per-frequency P2S model for each core type:

1. Train model on one core type and use the test set events from the other core-type as model inputs. This is shown to be a naive approach and is very inaccurate, particularly when trying to predict Cortex-A7 power with Cortex-A15 events (8000% average error across frequencies).
2. Match the average CPU power for benchmarks of one core type to the average event counts of the other core type to train and then test the data. This averaging is necessary, because we cannot exactly map the dynamic measurements power of one core to the events of the other due to uneven number of sample points.

The cross prediction model for the Cortex-A15 CPU power, trained and tested using Cortex-A7 events, comes really close to the P2S per-frequency model, both achieving 8% average prediction error. The trained cross-prediction model for the Cortex-A7 achieves a respectable accurate of 9.26%. Fig. 8 shows a per-frequency breakdown of the accuracy of the cross-prediction models.

These results indicate that it is possible to use the methodology to develop accurate runtime cross-prediction models for both core types of the big.LITTLE platform that could be used to guide an energy-aware task scheduler.

## VII. CONCLUSIONS

This paper has presented a novel hybrid approach for run-time power prediction and modelling in heterogeneous single ISA big.LITTLE processors. We evaluate our methodology by comparing against related work and our models achieve better accuracy - average 8% prediction error on the ARM Cortex-A15 and 5% on the ARM Cortex-A7. We have extended our models to include cross-prediction capabilities and aim to integrate them in a custom real-time scheduler, focused on optimising image processing/video streaming applications on mobile platforms. The ODROID XU3 platform is ideal for developing and testing initial work, but we also want to test the methodology on popular smartphones to how it would behave in a system witch strict thermal constraints. Our work could also be extended to other markets which feature heterogeneous architectures, particularly the server and HPC markets.

All our work is open source and can be viewed and downloaded at https://github.com/kranik/ARMPM

### ACKNOWLEDGEMENTS

This work is supported by ARM Research funding, through an EPSRC iCASE studentship, and the University of Bristol.

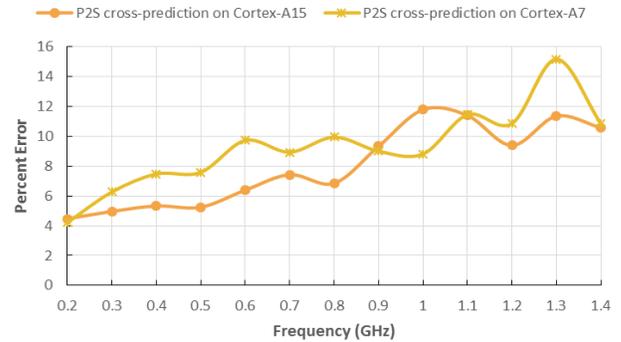

Figure 8. Inter-core power cross-prediction using P2S